\documentstyle[12pt]{article}
\newif\iffigs\figsfalse                                                  
\figstrue                                                                
\newif\ifbbB\bbBfalse                                                    
\bbBtrue                                                                 
\newif\ifdraft\draftfalse                                                

\title{}

\def\draftnote#1
  {\ifdraft 
    \hspace{1mm}\newline\noindent\begin{tabular}[t]{|p{14cm}|}
     \hline {\bf DRAFT NOTE}\\ #1 \\ \hline
    \end{tabular}\newline\noindent
   \fi}

\def\multdn{$\downarrow\downarrow\downarrow\downarrow\downarrow$}
\def\begindraft%
 {\newline\noindent\begin{tabular}[t]{|c|}
   \hline {\bf DRAFT NOTES}\\ 
   \multdn\multdn\multdn\multdn\multdn\multdn
   \multdn\multdn\multdn\multdn\multdn\multdn \\ \hline
  \end{tabular}\newline\noindent}

\def\multup{$\uparrow\uparrow\uparrow\uparrow\uparrow$}
\def\enddraft%
 {\newline\noindent\begin{tabular}[t]{|c|}
   \hline \multup\multup\multup\multup\multup\multup
          \multup\multup\multup\multup\multup\multup \\
   {\bf DRAFT NOTES}\\ \hline
  \end{tabular}\newline\noindent} 

\newcommand{\newsection}[1]{
 \vspace{10mm} \pagebreak[3]
 \addtocounter{section}{1}
 \setcounter{subsection}{0}
 \setcounter{paragraph}{0}
 \setcounter{equation}{0}
 \setcounter{figure}{0}
 \setcounter{table}{0}
 \addcontentsline{toc}{section}{\protect\numberline{\arabic{section}}{#1}}
 \begin{flushleft}
  {\large\bf \thesection. #1}
 \end{flushleft}
 \nopagebreak}

\def\al{\alpha}
\def\bt{\beta}
\def\gm{\gamma}                
\def\dl{\delta}                \def\Dl{\Delta}
\def\ep{\epsilon}

               \def\Lm{\Lambda}

\def\th{\theta}               
\def\vph{\varphi}

\def\Oc{\mbox{\protect$\cal O$}}

\ifbbB
 \font\blackboard=msbm10 
 \font\blackboards=msbm7 \font\blackboardss=msbm5
 \newfam\black \textfont\black=\blackboard
 \scriptfont\black=\blackboards \scriptscriptfont\black=\blackboardss
 \def\Bbb#1{{\fam\black\relax#1}}
\else
 \def\Bbb{\bf}
\fi
\def\CC{\Bbb{C}}

\def\ZZ{\Bbb{Z}}

\def\cl{{\rm cl}}
\def\diag{{\rm diag}}
\def\im{{\rm Im}}
\def\max{{\rm max}}

\def\Pf{{\rm Pf}}

\def\tr{{\rm tr}}
\def\eff{{\rm eff}}

\def\pt{\partial}
\def\goto{\rightarrow}

\def\half{{\scriptstyle {1 \over 2}}}
\def\vev#1{\left\langle#1\right\rangle}

\def\hs{\hspace{5mm}}
\def\hsc{\hspace{5mm},\hspace{5mm}}
\def\ie{{\em i.e.\ }}
\def\eg{{\em e.g.\ }}
\def\nl{\newline}

\def\beq{\begin{equation}}
\def\eeq{\end{equation}}

\def\Wtree{W_{\rm tree}}
\begin{document}



\begin{titlepage}

\ifdraft \fbox{\bf !!!!!!!!!!!!!! DRAFT VERSION !!!!!!!!!!!!!!} \fi

\begin{flushright}
RI-12-96\\
hep-th/9701045\\[5mm]
\ifdraft
 \count255=\time 
 \divide\count255 by 60
 \xdef\hourmin{\number\count255}
 \multiply\count255 by-60
 \advance\count255 by\time
 \xdef\hourmin{\hourmin:\ifnum\count255<10 0\fi\the\count255}
 \number\day/\number\month/\number\year\ \ \hourmin
\\[15mm]\fi
\end{flushright}

\begin{center}
\Large
{\bf The Coulomb Phase in $N=1$ Gauge Theories \\
With a LG-Type Superpotential}
\\[10mm]
\large
Amit Giveon, Oskar Pelc
\normalsize and \large
Eliezer Rabinovici
\normalsize
\\[5mm]
{\em Racah Institute of Physics, The Hebrew University\\
  Jerusalem, 91904, Israel}\\
E-mail: giveon@vms.huji.ac.il, oskar@shum.cc.huji.ac.il,\\
eliezer@vms.huji.ac.il
\\[15mm]
\end{center}

\begin{abstract}
We consider $N=1$ supersymmetric gauge theories with a simple classical gauge
group, one adjoint $\Phi$, $N_f$ pairs
($Q_i,\tilde{Q_i}$) of (fundamental, anti-fundamental)
and a tree-level superpotential with terms of the Landau-Ginzburg
form $\tilde{Q}_i\Phi^lQ_j$.
The quantum moduli space of these models includes a Coulomb branch. We find
hyperelliptic curves that encode the low energy effective gauge coupling for
the groups $SO(N_c)$ and $USp(N_c)$ (the corresponding curve for $SU(N_c)$ is
already known). As a consistency check, we derive the sub-space of some 
vacua with massless dyons via confining phase superpotentials. 
We also discuss the existence and nature of the non-trivial superconformal 
points appearing when singularities merge in the Coulomb branch. 
\end{abstract}

\draftnote{PACS codes: 11.15.-q, 11.15.Kc, 11.15.Tk, 11.30.Pb
\nl 11.15: Gauge FT; -q: general; Kc: Classical and Semiclassical 
  Techniques; Tk: Other Techniques; 11.30.Pb SuSy.
\nl Key words: Gauge Filed Theory, Supersymmetry, Coulomb Phase.}
\end{titlepage}

\ifdraft
 \pagestyle{myheadings}
 \markright{\fbox{\bf !!!!!!!!!!!!!! DRAFT VERSION !!!!!!!!!!!!!!}}
\fi

\flushbottom



\newsection{Introduction and Summary}

The introduction of supersymmetry into field-theoretical models 
extends the range of methods that can be used to analyze them. 
An important feature implied by supersymmetry is that certain 
quantities are restricted to depend holomorphically on their arguments
\cite{Seiberg-9309} (for a review, see for example \cite{IS-lec}).
This opens the way to the use of complex analysis and often leads to exact
determination of these quantities. One such quantity is the low energy 
effective superpotential. Its exact determination leads to the identification 
of the quantum moduli space of the model. Another such quantity is the 
low energy effective gauge coupling in the Coulomb branch of a gauge theory. 
The determination
of its dependence on the moduli leads to important qualitative as well as 
quantitative information about the low energy behavior: the spectrum of 
charged states, their masses, points of phase transition, non-trivial IR 
fixed points etc. .

In this work, we determine the effective gauge coupling in the Coulomb branch 
of a large family of $N=1$ models. One approach to this problem is to
consider directly the Coulomb branch, and combine the restrictions coming 
from symmetry and from various limits, to arrive at a unique expression. This
method led to the determination of the coupling for all the $N=2$ models with
a simple classical gauge group and fundamental matter 
\cite{SW}-\cite{AS9509}. Alternatively, one can approach the Coulomb branch
through the points where it meets the other (Higgs/confinement) branches.
These are singular points of the Coulomb branch (the effective coupling 
vanishes there) and, therefore, this approach provides information about the
singularity structure of the Coulomb branch which, because of the 
holomorphicity, determines the gauge coupling to a large extent. The singular 
points are determined by the low energy effective superpotential, which
in turn can be obtained in many cases by symmetry arguments. An efficient
method to obtain the superpotential is the ``integrating-in'' method
\cite{ILS,Intriligator}. This method was applied successfully to the $N=2$ 
models 
described above, to some of their $N=1$ generalizations \cite{IS9408}-%
\cite{Kitao}, and also to other models. Here we combine both approaches. 
Using the first approach, we determine the effective coupling and then
use the second approach as a consistency check in some examples. 

We consider an $N=1$ supersymmetric gauge theory with a simple classical gauge
group $G$, matter content of one adjoint $\Phi$ and $N_f$ pairs
($Q_i,\tilde{Q_i}$) of (fundamental, anti-fundamental)
and a tree level superpotential of the Landau-Ginzburg (LG) form
\beq\label{Wtree}
  \Wtree=\sum_l\tr(h_l Z_l) \hsc (Z_l)_{ij}=\tilde{Q}_i\Phi^lQ_j \hsc \eeq
where $Z_l$ are gauge invariant operators, and the sources $h_l$ are 
complex matrices%
\footnote{For the $SO(N_c)$ and $USp(N_c)$ groups, the fundamental and
anti-fundamental representations are equivalent, so there are $2N_f$
fundamentals and $Z_l, h_l$ are $(2N_f\times2N_f)$-dimensional.} 
\footnote{\label{N2} When $h_0$ is diagonal, $h_1=1$ and $h_l=0$ for $l>1$, 
  the model has $N=2$ supersymmetry.};
for $SU(N_c)$ such terms were considered in \cite{Kapustin}.
Note that the terms with $l>1$ are non-renormalizable and, therefore, the
model with such terms should be considered as a low energy effective model,
valid only for scales below $\sim h_l^{-\frac{1}{l-1}}$. Naively, one would
expect that such terms will be irrelevant in the IR but in fact this is not
true%
\footnote{The following argument is a generalization of an argument given
  for $G=SU(N)$ in \cite{Kapustin}.}.
Indeed, consider such a term as a perturbation of a scale invariant model.
The vanishing of the beta-function for the gauge coupling \cite{beta} implies
\[
  0=\bt_g \sim b+\sum k_i\gm_i
   =k_A(2+\gm_\Phi)+2N_fk_f[(2+\gm_Q)-3]
\]
(where $k_A$ and $k_f$ are the Dynkin indices of the adjoint and fundamental
representations%
\footnote{See Appendix A.}, 
$b=3k_A-\sum k_i=2k_A-2N_fk_f$ is minus the one loop coefficient,
and $\gm_\Phi,\gm_Q$ are the anomalous dimensions), therefore, at a fixed 
point
\draftnote{$\gm_Q=\gm_{\tilde{Q}}$ since $Q\leftrightarrow\tilde{Q}$ is a 
  symmetry; presumably non-anomalous.}
\beq
  \dim(\tilde{Q}\Phi^lQ)=(\gm_Q+2)+\half l(\gm_\phi+2)
  =3-\frac{2+\gm_\phi}{4N_fk_f}b' \hsc b'=2k_A-2lN_fk_f
\eeq
(note that $b'$ is related to $b$ by the replacement $N_f\goto lN_f$).
$(2+\gm_\phi)$ is the dimension of a gauge invariant operator tr$\Phi^2$ and, 
therefore, it is positive by unitarity. Consequently, the perturbation
$\tilde{Q}\Phi^lQ$ is relevant for $b'>0$, marginal for $b'=0$ and
irrelevant for $b'<0$.
In the following we will analyze only relevant perturbations $b'>0$ (some 
remarks about the other cases are presented at the end of this section). 
This implies 
that $b>0$, which means that the model is asymptotically free and at high
enough scales can be treated semi-classically.

Classically, the model with the superpotential (\ref{Wtree}) has a moduli
space of vacua with Coulomb ($Q=\tilde{Q}=0$) and Higgs branches. 
The Coulomb branch is parameterized by (gauge invariant functions of) the
vev $\vev{\Phi}$ of $\Phi$.
$\vev{\Phi}$ generically breaks the gauge symmetry to $U(1)^r$, where $r$ is
the rank of the gauge group. It also contributes to the mass of the quarks
$Q,\tilde{Q}$ through the superpotential (\ref{Wtree}).

As to the parameterization of the Coulomb branch, the D-term equations
imply that $\Phi$ can be chosen, by a color
rotation, to lie in the (complexified) Cartan sub-algebra of the gauge group.
Therefore, given a parameterization
$\{\phi_a\}$ of the Cartan sub-algebra, it is convenient to characterize
a point in the moduli space by functions of $\{\phi_a\}$, invariant under
the residual gauge freedom (the Weyl group).

The Coulomb branch vacua ($Q=\tilde{Q}=0$) survive also in the
quantum theory%
\footnote{\label{Q-Coul} As discussed in \cite{Kapustin} for $G=SU(N_c)$, the 
  theory has an anomaly-free R-symmetry with $R_\Phi=0$ and
  $R_Q=R_{\tilde{Q}}=1$. Therefore, any dynamically generated superpotential
  is quadratic in $Q,\tilde{Q}$ and cannot lift the Coulomb branch.}.
However, they can be modified by quantum corrections. In the weak-coupling
region (where the gauge invariance is broken at high scale) the classical
parameterization is valid, therefore, we will adopt for the whole quantum
moduli space the same coordinates $\{\phi_a\}$ as in the classical moduli
space, with the understanding that {\em a priori} they have the classical
meaning (\ie the classical relation to $\vev{\Phi}$) only in the
semi-classical region.
\draftnote{There is an implicit assumption here that such a parametrization 
    exists (no topological obstructions).}

At a generic point in the Coulomb moduli space, the low energy effective
theory is a $U(1)^r$ gauge
theory with no massless charged matter, and the kinetic term in the effective
action is of the form
\beq \frac{1}{16\pi}\im\int d^4xd^2\th\tau_\eff^{\al\bt}W_\al W_\bt. \eeq
The effective gauge coupling matrix $\tau_\eff$ is a function of the moduli,
bare parameters and the microscopical coupling constant (represented, in the
asymptotically free cases, by the dynamically generated scale $\Lm$).
Supersymmetry implies that it
is holomorphic (for the {\em Wilsonian} effective action), electric-magnetic
duality implies that it is not single valued but rather an $Sp(2r,\ZZ)$ section
and unitarity implies that it is positive-definite (for a review, see for
example \cite{IS-lec}). 
All this suggests  that
$\tau_\eff$ may be identified as the period matrix of a genus $r$
Riemann surface, parameterized holomorphically by $\vev{\Phi}$, $h_l$ and
$\Lm$. Moreover, as in \cite{KLYT}-\cite{AS9509}, we will assume
\draftnote{What can be deduced {\em a posteriory} from the uniqueness 
  of the Riemann-Hilbert problem ?}
that this surface is a hyper-elliptic curve, \ie, described by 
\beq\label{y2K}
  y^2=K(x;\phi_a,h_l,\Lm^b)=k\prod_{l=1}^{2r+\dl}(x-x_l)
  \hsc \dl=1\mbox{ or }2\hs
\eeq
and that $K$ depends polynomially also on $\phi_a$, $h_l$ and the instanton
factor $\Lm^b$. For the group $SU(N_c)$, such a description was found in 
\cite{Kapustin}. We will look for the corresponding description for the 
other classical groups $SO(N_c)$ and $USp(N_c)$.

Before proceeding with the derivation of the curves, let us summarize the 
results: 
\begin{itemize}
\item $SU(N_c)$, $N_c=r+1$: ($b=2N_c-N_f$)
\beq\label{SU-curve}
  y^2=P(x)^2-\Lm^b H(x) \hsc 
\eeq
\[ P(x)=\prod_{a=1}^{r+1}(x-\phi_a) \hsc H(x)=\det h(x) \hs; \]

\item$SO(N_c)$, $N_c=2r+\ep$, $\ep=0,1$: ($b=2(N_c-2-N_f)$)
\beq\label{SO-curve} 
  y^2=xP(x)^2+\Lm^b x^{3-\ep}H(x) \hsc 
\eeq
\[ P(x)=\prod_{a=1}^r (x-\phi_a^2) \hsc H(x)=\det(h(\sqrt{x})) \hs; \]

\item $USp(N_c)$, $N_c=2r$: ($b=N_c+2-N_f$)
\beq\label{Sp-curve} 
  xy^2=(xP(x)+\Lm^b\Pf(h_0))^2-\Lm^{2b} H(x) \hsc 
\eeq
\[P(x)=\prod_{a=1}^r (x-\phi_a^2) \hsc H(x)=\det(h(\sqrt{x})) \hs. \]

\end{itemize}
In all cases, we denote 
\beq h(t)=\sum_{l=0}^{l_{\max}} h_l t^l. \eeq
For $SO(N_c)$ and $USp(N_c)$, because the fundamental and 
anti-fundamental representations are equivalent, $h_l$ is $(2N_f\times2N_f)$%
-dimensional. For $SU(2)$, being identical to $USp(2)$, a distinction 
between fundamentals and anti-fundamentals is also artificial. Nevertheless,
the $SU(2)$ curve in eq. (\ref{SU-curve}) is for the special case, where
such a distinction is made, \ie, the $2N_f$ quarks are divided into two subsets
 -- fundamentals and ``anti-fundamentals'' -- and $h_l$ (an 
$(N_f\times N_f)$-dimensional matrix) couples only quarks belonging to 
different subsets. The $SU(2)$ curve in eq. (\ref{Sp-curve}) is for the 
general case. 

For the range of parameters in
which the model has $N=2$ supersymmetry (see footnote \ref{N2}) our results
coincide with the previous results \cite{APS9505,AS9509}.
Note that these curves are indeed of the form (\ref{y2K}): in all cases, 
$H(x)$ is a {\em polynomial} in $x$ 
(for the last two cases, this is due to the symmetry
properties of $h$) and in the $USp(N_c)$ case the right-hand side is divisible
by $x$. The degree of $y^2$ is either $2r+2$ (in eq. (\ref{SU-curve})) or
$2r+1$ (in eqs. (\ref{SO-curve}), (\ref{Sp-curve})) and, therefore, the curve
has the correct genus. 

These curves were originally derived for $b'>0$, where
$b'=2k_A-2l_\max N_fk_f$. However, 
{\em a posteriori}, by breaking the gauge symmetry to a subgroup of the same
type (as is done in the subsequent sections), one can attempt to derive the 
curves for $b'\le0$. In these cases, there exists, in particular, a term of 
the form $\Lm^bG(h)$ ($G$ being a polynomial in the components of $h_l$)%
\footnote{$G(h)$ 
  is obtained by expanding $H(x)$ in powers of $x$ and taking the
  appropriate coefficient, which exists if the degree of $H(x)$ is high 
  enough, and this happens exactly when $b'\le0$.} 
which is invariant under all symmetries and, therefore, it is harder to
determine its appearance in the curve 
(in the process of symmetry breaking, it can enter
through the matching conditions). Recall, however, that for%
\footnote{The case $l_\max=1$ is special, but it is essentially the $N=2$
case.}
$l_\max>1$, the 
model is non-renormalizable, valid only at a sufficiently low scale, so $\Lm$
(and, therefore, also $\Lm^bG$) is restricted to be sufficiently small, and
one may hope that in this range the dependence on $\Lm^bG$ is weak enough
so that it can be ignored. When this is true, the curves have the above form 
also for $b'\le0$ (where $\Lm^0$, in the $b=0$ cases, is understood to
represent an appropriate modular function of the microscopic gauge coupling).

For $b'<0$, there is another complication: the degree of $H(x)$ is so high that
the curve has genus greater then required (there are more then $2r+2$ branch
points). In the $N=2$ case this happens
when the model is IR free ($b<0$) and, therefore, valid at scales small 
compared
to $\Lm$. As was discussed in \cite{APS9603}, this also implies that the
curve is valid only for $x<\Lm$ ($x<\Lm^2$ for $SO(N_c)$ and $USp(N_c)$) and 
in this region there are exactly the right number of branch points. For
$l_\max>1$, there also exists a restriction to low scales (even when $b\ge0$), 
coming from non-renormalizability so, presumably, also in this case the
extra branch points are outside the range of validity of the curve.

The outline of the next sections is as follows.
In section 2, we derive the curves for $SO(N_c)$ and $USp(N_c)$,
using the first approach described above. In section 3, we derive the 
singularity equations for the curves of $SU(N_c)$ and $SO(N_c)$, approaching
from the Higgs-confinement branch, and compare them with the known curves. 
In section 4, we discuss the existence and nature of points in the
Coulomb branch that correspond to non-trivial superconformal theories.
Finally, in two appendices, we present some details.

\newsection{Determination of the Curves}

We assume that $\tau_\eff$ is the period matrix of a genus $r$ 
hyper-elliptic curve of the form (\ref{y2K})
and that $K$ depends polynomially also on $\phi_a$, $h_l$ and the one instanton
factor $\Lm^b$. We will consider the expansion of $K$ in powers of
$\Lm^b$:
\beq\label{Kal} 
  K=\sum_{\al=0}^{\al_{\rm max}} \Lm^{b\al}K_\al(x;\phi_a,h_l) \hs.
\eeq
This obviously depends on the coordinates $(x,y)$ chosen to describe the curve.
The coordinates will be chosen to coincide with those of the $N_f=0$ model.

The curve should be invariant under the (gauge and global) symmetries of
the model (since $\tau_\eff$ is). This includes all symmetries of the
classical model without a superpotential%
\footnote{Explicit breaking by the superpotential
  is accounted for by transforming also the parameters $h_l$ and anomalies are
  compensated by transforming the instanton factor $\Lm^b$.}.
Three $U(1)$ symmetries are particularly useful in the discussion. Adapting 
the notation
\[ \begin{array}{|c||c|c|c|}\hline
         & \th & Q & \phi \\ \hline
  Q_\th  &   1 & 0 &    0 \\
  Q_Q    &   0 & 1 &    0 \\
  Q_\Phi &   0 & 0 &    1 \\ \hline
\end{array}\]
($\th$ is the fermionic coordinate of superspace), we define
\[ R'=Q_\th+Q_Q \]
\beq\label{U1} R=Q_\th+2Q_\Phi \eeq
\[ A=Q_Q \]
($R'$ is the symmetry used in footnote \ref{Q-Coul}). Unlike $R'$ which is a 
symmetry of the full model, both $R$ and $A$ are anomalous and are also broken 
explicitly by the superpotential. 
This means that $\Lm^b$ and $h_l$ have non-trivial 
$R$ and $A$ charges and, therefore, their appearance in the curve is 
restricted by these symmetries. Note also that for all the parameters appearing
in the curve, $\half R$ coincides with their mass dimension.

In the following we consider the classical groups case by case.

\subsection{$USp(N_c)$, $N_c=2r$}

\subsubsection{Generalities}

The fundamental ($N_c$ dimensional) representation of $USp(N_c)$ consists of
unitary matrices $U$ that obey
\beq\label{JU} U^TJU=J \hsc
  J=\left(\begin{array}{cc}0 & 1 \\ -1 & 0 \end{array}\right)\otimes I_r \hsc
\eeq
where $I_r$ is the $r$-dimensional identity matrix. This means that the
fundamental representation is pseudo-real (\ie, $JQ_i$ transforms under the
anti-fundamental representation), so the superpotential (\ref{Wtree}) can
take the more general form
\beq\label{Sp-Wtree}
  \Wtree=\sum_l\tr(h_l Z_l) \hsc (Z_l)_{ij}=Q_i^T J\Phi^lQ_j \hsc \eeq
where $h_l$ are now ($2N_f\times2N_f$)-dimensional. The relation (\ref{JU})
also implies that $J\Phi$ is a symmetric matrix, therefore,
\beq\label{Sp-hl} Z_l^T=(-1)^{l+1}Z_l \hsc h_l^T=(-1)^{l+1}h_l\hs. \eeq
The vev $\vev{\Phi}$, after gauge rotation, takes the form
\beq\label{Sp-Phi} \vev{\Phi}=
  \left(\begin{array}{cc}1 & 0 \\ 0 & -1 \end{array}\right)
  \otimes\diag\{\phi_a\} \hsc \phi_a\in\CC.
\eeq
The residual gauge freedom is generated by permutations and flips of sign
of the $\{\phi_a\}$, so the gauge invariant combinations are $\{s_{2k}\}$,
defined by the generating function
\beq\label{Sp-P}
  P(x;\phi_a)=\sum_{k=0}^r s_{2k} x^{r-k}=\prod_{a=1}^r(x-\phi_a^2).
\eeq
The instanton factor is $\Lm^b$, where $b=2r+2-N_f$. 

\subsubsection{Determining $K_0$}

One can give the $Q$'s a large mass and integrate them out:
\beq\label{m-inf}
\Pf(h_0)\goto\infty \hsc \Lm_0^{2r+2}=\Pf(h_0)\Lm^b \hs,
\eeq
obtaining the $N_f=0$ model (which has $N=2$ supersymmetry), for which the
curve is known \cite{AS9509}:
\draftnote{{\em A priori} there are also other possibilities: $(xP)^2$
  or $x^2P$ multiplied by some other polynomial in $x$ and $s_{2k}$ with the
  right dimension ($R$-charge). The first choice was ruled out in
  \cite{AS9509} by the (unnecessary) requirement of uniformity in $r$,
  knowing the curve for $r=1$.} 
\beq\label{Sp-K0} y^2=xP^2+O(\Lm^b). \eeq
We {\em choose} the coordinates $(x,y)$ of the curve
for the $N_f\neq0$ model to be those of the $N_f=0$ model
({\em i.e.}, in the above
limit we should obtain the $N_f=0$ curve with the same $(x,y)$). This fixes 
the charges of $x$ and $y$ under the symmetries of the model.
The charges of the various quantities appearing in the curve under the
$R$ and $A$ symmetries are%
\footnote{$R'(\Phi)=R'(h_l)=R'(\Lm^b)=0$ while $R'(W_\eff)=2$, therefore, 
as stated in the introduction, an effective superpotential cannot be 
generated in the Coulomb phase, where $Q=\tilde{Q}=0$, and
the classical moduli space in the Coulomb branch is not lifted quantum 
mechanically.}:
\beq\label{Sp-U1}\begin{array}{|c||c|c|c||c|c||c|c|}\hline
    & \th & Q & \phi & \Lm^b &    h_l & x &         y \\ \hline
  R &   1 & 0 &    2 &      2b & 2(1-l) & 4 & 2(2r+1) \\
  A &   0 & 1 &    0 &      2N_f &     -2 & 0 &         0 \\ \hline
\end{array}\eeq
Using the $A$-charge we now observe that $K_0$ is independent of $N_f$ and,
therefore, given by the $N_f=0$ curve. Indeed, $K_\alpha$ must have vanishing 
$A$-charge (since $x$ and $y$ do). $K_0$ is, by definition, independent of 
$\Lm$ and, therefore (by the presumed polynomial dependence), also independent
 of $h_l$. In particular, it is invariant under the limit (\ref{m-inf}).

\subsubsection{$USp(2\hat{r})\goto USp(2r)$}

In the curve for $USp(2\hat{r})$ we set
\beq \hat{\phi}_a=\left\{\begin{array}{lc}
    \phi_a    & 1\leq a\leq r  \\
    \phi'_a+M & r<a\leq\hat{r}
  \end{array}\right.
\eeq
and take $M\goto\infty$. Classically, for $\phi_a=0$ (and generic $\phi'_a$)
the gauge symmetry $USp(2\hat{r})$ is broken to 
$USp(2r)\times U(1)^{\hat{r}-r}$ and
$\phi_a$ parameterize the Coulomb moduli space of a $USp(2r)$ model. Therefore,
we expect to obtain, in the above limit, the curve of the $USp(2r)$ model.
We must identify the details of this model:
the fundamentals $\hat{Q}_i$ of $USp(2\hat{r})$ decompose, each, to one
fundamental $Q_i$ of $USp(2r)$ (consisting of the first $2r$ components of
$\hat{Q}_i$) and $USp(2r)$-singlets; the light components of the adjoint
$\hat{\Phi}$ are the components of an adjoint of $USp(2r)$ (the upper 
$2r\times2r$ block), so the $USp(2r)$ model has the same matter content as the
$USp(2\hat{r})$ one. The tree level superpotential of $USp(2\hat{r})$
reduces to
\[ \hat{W}_{\rm tree}=\sum_l\tr\hat{h}_l\hat{Q}^T\hat{J}\hat{\Phi}^l\hat{Q}
                     \goto \sum_l\tr\hat{h}_l Q^T J\Phi^l Q,  \]
which implies $\hat{h}_l=h_l$. Finally, the matching of scales is%
\footnote{This is true at least in the semiclassical region. If the $USp(2r)$
  model is asymptotically free, then the limit $M\goto\infty$ corresponds to
  vanishing coupling at the matching point, so in this case the relation
  is exact.}
\beq \hat{\Lm}^{\hat{b}}=\Lm^bM^{2n'} \hsc n'=\hat{r}-r    \eeq
(up to a scheme dependent multiplicative constant that can be absorbed in 
$\Lm$). The curve for the $USp(2\hat{r})$ model is
\[ \hat{y}^2=\hat{x}\hat{P}^2+\hat{\Lm}^{\hat{b}}\hat{K}_1
                             +\hat{\Lm}^{2\hat{b}}\hat{K}_2
                             +\Oc(\hat{\Lm}^{3\hat{b}})  \]
with
\[ \hat{P}(\hat{x})=\prod_1^r(\hat{x}-\phi_a^2)
                    \prod_{r+1}^{\hat{r}}(\hat{x}-(\phi'_a+M)^2). \]
To obtain a curve of the form (\ref{Sp-K0}) in the limit $M\goto\infty$,
one must identify $\hat{x}=x$, $\hat{y}=yM^{2n'}$, so for $x\ll M^2$ the
curve is
\beq y^2=xP^2+\Lm^b\hat{K}_1/M^{2n'}
            +\Lm^{2b}\hat{K}_2+\Oc(M^{2n'})  \eeq
and $\hat{K}_\al$ depends on $M$ polynomially, through 
$\{\hat{\phi}_a\}=\{\phi_a\}\cup\{\phi'_a+M\}$.
Since the limit $M\goto\infty$ is finite, we get the following restriction 
on the general form (\ref{Kal}):
\begin{itemize}
  \item $\hat{K}_\al=0$ for $\al>2$;
  \item $\hat{K}_2$ is independent of $\hat{\phi}_a$;

     (it is independent of $\phi'_a$ and, since the dependence on 
      $\hat{\phi}_a$ is only through symmetric functions of them, there 
      cannot be a dependence on $\phi_a$ either).
\end{itemize}
These restriction are carried over also to the $USp(2r)$ curve, so we 
conclude that the curve is of the form
\beq y^2=xP(x;\phi_a)^2+\Lm^bK_1(x;\phi_a,h_l)+\Lm^{2b}K_2(x;h_l). \eeq

\subsubsection{The Semi-Classical Quark Singularities}

In the semi-classical region (weak coupling), we can substitute
$\Phi\goto\vev{\Phi}$ from (\ref{Sp-Phi}) in the superpotential
(\ref{Sp-Wtree}) and obtain the effective mass for the fundamentals
\beq\label{Sp-Meff}
  \Wtree\goto\sum_a Q_a^T\left(\begin{array}{cc}
  0 & h^T(-\phi_a) \\ -h^T(\phi_a) & 0 \end{array}\right)Q_a \hsc
  h(t)=\sum_l h_l t^l \hsc
\eeq
where $Q_i$ is partitioned to 2-component fields $Q_{ia}$ (according to the
block structure of $\vev{\Phi}$) and the summation over flavor indices is
implicit. Eq. (\ref{Sp-hl}) implies that
\beq\label{Sp-h} h(-t)=-h(t)^T, \eeq
therefore,
\beq\label{Sp-m-eff} \Wtree\goto Q^T m_\eff Q \hsc
  m_\eff=-\diag\left\{\left(\begin{array}{cc}
  0 & h(\phi_a) \\ h^T(\phi_a) & 0 \end{array}\right)\right\}.
\eeq
Whenever det($m_\eff)=0$, there is a massless d.o.f. charged under (at least)
one of the $U(1)$ factors of the residual gauge freedom. The effective coupling
of this $U(1)$ vanishes in the IR, which means that the curve must be singular
in such cases and this is characterized by the vanishing of the discriminant
$\Dl_K$ of the curve. Since (\ref{Sp-m-eff}) is a classical expression, we
should compare it with the $\Lm\goto0$ limit of $\Dl_K$, but we cannot just
substitute $\Lm=0$ since in this case $\Dl_K$ vanishes identically. What
we need is the leading term, \ie:
\beq \Dl_K=\ep^\al\Dl_K^0+o(\ep^\al) \hsc \ep=\Lm^b \hsc \eeq
where $\Dl_K^0$ is a polynomial%
\footnote{Recall that the discriminant $\Dl_K$ is a polynomial in the
coefficients of $K$ and, therefore, it is a polynomial in $\phi_a$, $h_l$ and
$\ep$.}
in $\phi_a$ and $h_l$, independent of $\ep$ but non-trivial. The zeros of
$\Dl_K^0$ are the $\Lm\goto0$ limits of the singular points in the moduli
space and, therefore, $\Dl_K^0$ should vanish whenever det$(m_\eff)=0$.
Since
\[ \det(m_{\eff})=\prod_a H(\phi_a^2)^2 \hsc H(x)=\det(h(\sqrt{x})) \]
(note that, because of (\ref{Sp-h}), $\det(h(t))=\det(-h(t))$, hence $H$ is 
an {\em even} polynomial of $\sqrt{x}$ and, therefore, a {\em polynomial} in 
$x$), the requirement that we obtain is
\beq\label{Sp-DlK-H} \Dl_K^0|H(\phi_a^2)\;\forall a \eeq
(\ie, $H(\phi_a^2)$, as a polynomial in $\phi_a^2$, is a factor in $\Dl_K^0$).

In the calculation of $\Dl_K$, it is enough to consider generic values of
the parameters (the rest is determined by continuity). We therefore assume
$\Dl_{xP}\neq0$ (which means that all the $\phi_a$'s are different from
each other and from $0$) and $H(0)\neq0$ (which means that all the quarks
have a non-trivial bare mass). The calculation is done in Appendix B and here 
we use the results. First we eliminate a few possibilities:
\begin{itemize}
  \item{$K_1|_{x=\phi_a^2}\not\equiv0$}

    (a non-trivial polynomial in $\phi_a^2$ and $h_l$)%
\footnote{\label{sym} Note that if this is true for one $a$ it is true for all
      of them, by symmetry.}.

    Eqs. (\ref{Dl-xPP}) and (\ref{Kp1-xPP}) give
    \[ \Dl_K\sim(\Dl_P)^2\prod_a(x^3K_1(x))_{x=\phi_a^2}+\Oc(\ep) \]
    (as in the Appendix, ``$\sim$'' means equality up to a numerical
    constant), which means that
    \beq \Dl_K^0\sim(\Dl_P)^2\prod_a(x^3K_1(x))_{x=\phi_a^2}. \eeq
    Eq. (\ref{Sp-DlK-H}) now implies that $x^3K_1|H$ (as polynomials in $x$),
    but the $A$-charge of $\Lm^b H$ is $-2N_f$, which would imply that
    $\Lm^b K_1$ is not $A$-invariant. Therefore this possibility is ruled out.

  \item{$K_1\equiv0$}

    Using eqs. (\ref{Dl-xPP}) and (\ref{Kp1-xPP}) again, this time with
    $\ep\goto\ep^2$, and $K_1\goto K_2$, one obtains
    \[ \Dl_K\sim(\Dl_P)^2\prod_a(x^3K_2(x))_{x=\phi_a^2}+\Oc(\ep). \]
    $K_2(x)$ is a non-trivial polynomial (since otherwise the curve would be
    singular for any $\phi_a$ and $h_l$) and does not depend on $\phi_a$ (as
    shown in the previous subsection), therefore, for generic $\phi_a$'s,
    $K_2(\phi_a^2)$ does not vanish and this means that
    \beq \Dl_K^0\sim(\Dl_P)^2\prod_a(x^3K_2(x))_{x=\phi_a^2}. \eeq
    Eq.(\ref{Sp-DlK-H}) now implies that $x^3K_2|H$ (as polynomials in $x$).
    $A$-charge conservation implies that $x^3K_2/H$ is independent of $h_l$ and
    by $R$-charge conservation we obtain $K_2=H/x$, which is impossible,
    since $H(0)=\det(h_0)\neq0$.
\end{itemize}

The only other possibility is that $K_1$ is non-trivial but
$K_1|_{x=\phi_a^2}$ is. This means that $K_1=2PQ$, where $Q(x;\phi_a^2,h_l)$
is non-trivial. Eq. (\ref{Dl-xPP}) and (\ref{Kp1-xPP}) gives, again,
\[ \Dl_K\sim(\Dl_P)^2 P(0)^3\prod_a K(p'_a) \hsc \] 
but this time $K_1(p_a)=0$, so we must use eq. (\ref{Kp0-xPP}), which gives
\beq K(p'_a)\sim\left(\frac{\tilde{H}}{x}\right)_{x=\phi_a^2}+o(\ep^2)
   \hsc \tilde{H}=Q^2-xK_2. \eeq
Next we show that $\tilde{H}(\phi_a^2)$ is a non-trivial polynomial in
$\phi_b^2$. By contradiction, assume that%
\footnote{See footnote \ref{sym}.}
$\tilde{H}(\phi_a^2)=0$. Then
\[ Q(x;\phi_b^2)|_{x=\phi_a^2}=\phi_a^2 K_2(\phi_a^2) \hsc \]
which implies that $Q(x)$ is independent of $\phi_b^2$ and, moreover,
\[ Q=x\tilde{Q} \hsc K_2=x\tilde{Q}^2. \]
But this leads to
\[ y^2=x(P+\ep\tilde{Q})^2 \hsc \]
which is singular for any $\phi_a^2$. We therefore conclude that
$\tilde{H}(\phi_a^2)$ is indeed non-trivial, which implies that
\beq \Dl_K^0\sim(\Dl_P)^2\prod_a (x^2\tilde{H}(x))_{x=\phi_a^2}. \eeq
Eq.(\ref{Sp-DlK-H}) now implies, since $H(0)\neq0$, that
$\tilde{H}|H$ (as polynomials in $x$) and the conservation of the $A$ and $R$
charges implies that they are proportional. The proportionality constant can
be absorbed in $\ep$ and $Q$, leading to the curve
\beq xy^2=(xP+\Lm^b Q)^2-\Lm^{2b} H. \eeq
It remains to determine $Q(x)$. The relation $Q^2=xK_2+H$ implies that $Q$
is independent of $\phi_a^2$ (since the right hand side is)
and $Q(0)=\Pf(h_0)$ (since $Q^2-H$ is divisible by $x$). Then, symmetry 
($SU(2N_f)$, $A$ and $R$) restrict $Q$ to the following form
\[ Q(x;h_l)=\Pf(\sum_ia_ih_{2i}x^i) \hs (a_0=1), \]
where $a_i$ are numerical coefficients. This implies that the degree of 
$Q(x)$ is at most $\half l_\max N_f$ and, since we restrict ourselves to 
relevant perturbations $l_\max N_f<2r+2$, deg$Q$ is smaller then deg$(xP)$. 
This means that, expanding $Q$ in powers of $x$, all the coefficients, except 
the constant term, can be absorbed in $P$, by redefining the gauge invariant 
moduli:
\beq s_{2k}\goto s_{2k}-\Lm^b q_{r-k+1}, \mbox{   where   } 
   Q(x)=\sum_k q_k(h_{2l}) x^k  .      \eeq
To summarize, we have determined the curve to be (\ref{Sp-curve}).

\subsection{$SO(N_c)$, $N_c=2r+\ep$, $\ep=0,1$}

\subsubsection{Generalities}

The fundamental ($N_c$ dimensional) representation of $SO(N_c)$ is real, 
so the superpotential (\ref{Wtree}) can take the more general form
\beq\label{SO-Wtree}
  \Wtree=\sum_l\tr(h_l Z_l) \hsc (Z_l)_{ij}=Q_i^T \Phi^lQ_j \hsc \eeq
where $h_l$ are now ($2N_f\times2N_f$)-dimensional. $\Phi$ is an
anti-symmetric matrix, therefore,
\beq\label{SO-hl} Z_l^T=(-1)^l Z_l \hsc h_l^T=(-1)^l h_l. \eeq
The vev $\vev{\Phi}$, after gauge rotation, takes the form
\beq\label{SO-Phi} \vev{\Phi}=
  i\left(\begin{array}{cc}0 & -1 \\ 1 & 0 \end{array}\right)
  \otimes\diag\{\phi_a\} \hsc \phi_a\in\CC
\eeq
(for odd $N_c$, there is an additional row and column of zeros).
The residual gauge freedom is generated by permutations and flips of sign
of the $\{\phi_a\}$, so the gauge invariant combinations are $\{s_{2k}\}$,
defined, as for $USp(N_c)$, by%
\footnote{More precisely, for even $N_c$, the flips of signs come in pairs,
  and this fact is reflected by the existence of another invariant 
  $\Pf(\Phi)$. However, using symmetry arguments, one can show \cite{AS9509}
  that the curve can depend only on $\Pf(\Phi)^2$, which is $s_{2r}$.}
\beq\label{SO-P}
  P(x;\phi_a)=\sum_{k=0}^r s_{2k} x^{r-k}=\prod_{a=1}^r(x-\phi_a^2).
\eeq
The instanton factor is $\Lm^b$ with  $b=2(N_c-2-N_f)$. 
The determination of the curve uses arguments identical to the $USp(N_c)$ case,
so we will concentrate on the differences in the details.  

\subsubsection{Determining $K_0$}

$K_0$ is determined, by considering the $N_f=0$ model, to be \cite{AS9509}
\draftnote{{\em A priori} there are also other possibilities: $(xP)^2$ 
  or$x^2P$ multiplied by some other polynomial in $x$ and ${s_2k}$ with the
  right dimension ($R$-charge). The first choice was ruled out in
  \cite{AS9509} by the (unnecessary) requirement of uniformity in $r$,
  knowing the curve for $r=1$.}
\beq\label{SO-K0} y^2=xP^2+O(\Lm^b). \eeq
The charges of the various quantities appearing in the curve under the $R$ and
$A$ symmetries are:
\beq\label{SO-U1}\begin{array}{|c||c|c|c||c|c||c|c|}\hline
    & \th & Q & \phi & \Lm^b &    h_l & x &         y \\ \hline
  R &   1 & 0 &    2 &      2b & 2(1-l) & 4 & 2(2r+1) \\
  A &   0 & 1 &    0 &      4N_f &     -2 & 0 &         0 \\ \hline
\end{array}\eeq

\subsubsection{$SO(\hat{N}_c)\goto SO(N_c)$, $\hat{N}_c-N_c=2n'$}

In the curve for the $SO(\hat{N}_c)$ model we set
\beq \hat{\phi}_a=\left\{\begin{array}{lc}
    \phi_a    & 1\leq a\leq r  \\
    \phi'_a+M & r<a\leq\hat{r}
  \end{array}\right.
\eeq
so, in the limit $M\goto\infty$ we should obtain the curve of the $SO(N_c)$
model. As in $USp(N_c)$, the matter content is the same, $\hat{h}_l=h_l$ and 
the matching of scales is%
\footnote{When the $SO(N_c)$ model is asymptotically free, this relation
  is exact; see the footnote in the corresponding section for $USp(N_c)$.}
\beq \hat{\Lm}^{\hat{b}}=\Lm^bM^{4n'} \hsc n'=\hat{r}-r    \eeq
(up to a multiplicative constant that can be absorbed in $\Lm$).
The curve for the $SO(\hat{N}_c)$ model is
\[ \hat{y}^2=\hat{x}\hat{P}^2+\hat{\Lm}^{\hat{b}}\hat{K}_1
                             +\Oc(\hat{\Lm}^{2\hat{b}})  \]
with
\[ \hat{P}(\hat{x})=\prod_1^r(\hat{x}-\phi_a^2)
                    \prod_{r+1}^{\hat{r}}(\hat{x}-(\phi'_a+M)^2). \]
To obtain a curve of the form (\ref{SO-K0}) in the limit $M\goto\infty$,
one must identify $\hat{x}=x$, $\hat{y}=yM^{2n'}$, so for $x\ll M^2$ the
curve is
\[ y^2=xP^2+\Lm^b\hat{K}_1+\Oc(M^{4n'})  \]
and, since the limit $M\goto0$ is finite, the curve must be of the form
\beq y^2=xP(x;\phi_a)^2+\Lm^bK_1(x;h_l). \eeq
 
\subsubsection{The Semi-Classical Quark Singularities}

In the semi-classical region (weak coupling), we can substitute
$\Phi\goto\vev{\Phi}$ from (\ref{SO-Phi}) in the superpotential
(\ref{SO-Wtree}) to obtain the effective mass for the fundamentals. 
We diagonalize $\vev{\Phi}$ by a $U(N_c)$ rotation (which is a global
symmetry)%
\footnote{As in (\ref{SO-Phi}), we omit the additional trivial row and
  column for odd $N_c$.}
\beq\label{rot} \vev{\Phi}\goto U^\dagger \vev{\Phi} U=
   \left(\begin{array}{cc}-1 & 0 \\ 0 & 1 \end{array}\right)
  \otimes\diag\{\phi_a\} \hsc U=\frac{1}{\sqrt{2}}
   \left(\begin{array}{cc} 1 & 1 \\ -i & i \end{array}\right) \otimes I_r
\eeq
and obtain
\beq\label{SO-Meff}
  \Wtree\goto\sum_a Q_a^T\left(\begin{array}{cc}
  h^T(-\phi_a) & 0 \\ 0 & h^T(\phi_a) \end{array}\right)Q_a
  +{Q^{(2r+1)}}^T h_0^T Q^{(2r+1)} \hsc h(t)=\sum_l h_l t^l,
\eeq
where the $U$-transformed $Q_i$ is partitioned to 2-component fields $Q_{ia}$
(according to the block structure of $\vev{\Phi}$) and the summation over
flavor indices is implicit. $Q^{(2r+1)}$ (which appears only for odd $N_c$)
is not charged under the $U(1)^r$ residual gauge freedom and, therefore, has
no influence on the effective coupling, so we will ignore it in the following.
Eq. (\ref{SO-hl}) implies that
\beq\label{SO-h} h(-t)=h(t)^T, \eeq
therefore,
\beq\label{SO-m-eff} \Wtree\goto Q^T m_\eff Q \hsc
  m_\eff=\diag\left\{\left(\begin{array}{cc}
  h(\phi_a) & 0\\ 0 & h^T(\phi_a) \end{array}\right)\right\}.
\eeq
As for $USp(N_c)$, in the calculation of the discriminant $\Dl_K$, we 
consider the generic case
$\Dl_{xP}\neq0$ (which means that all the $\phi_a$'s are different from
each other and from $0$) and $H(0)\neq0$ (which means that all the quarks
have a non-trivial bare mass). $K_1(x)$ is a non-trivial polynomial in $x$
which does not depend on $\phi_a$, therefore, $K_1|_{x=\phi_a^2}$ is a 
non-trivial polynomial in $\phi_a^2$. This is the first case considered for
$USp(N_c)$ and it leads to the requirement $x^3K_1|H$ where 
$H(x)=\det(h(\sqrt{x}))$ (which is a polynomial in $x$). 
Now the $A$ and $R$ symmetries determine $K_1$ up to a constant, which can
be absorbed in $\Lm$. Therefore, the curve is determined to be 
(\ref{SO-curve}).

\subsection{$SU(N_c)$}

The vev $\vev{\Phi}$, after a gauge rotation, takes a diagonal form
\beq\label{SU-Phi}
  \vev{\Phi}=\diag\{\phi_a\} \hsc \phi_a\in\CC \hsc \sum_{a=1}^{N_c}\phi_a=0.
\eeq
The residual gauge freedom is generated by permutations, so the gauge
invariant combinations are $\{s_k\}$, defined by the generating function
\beq\label{SU-P} P(x;\phi_a)=\sum_{k=0}^{N_c}s_k x^{N_c-k}=
  \det(x-\vev{\Phi})=\prod_{a=1}^{N_c}(x-\phi_a).
\eeq
Using similar considerations as for the previous groups, one finds for this 
model the curve (\ref{SU-curve}).

\newsection{Vacua With Massless Dyons}
\label{Sec-Dyon}
The verification that the curve, conjectured to describe the Coulomb branch
of the moduli space, indeed gives the correct singularity structure of
the moduli space is a restrictive consistency check. The purpose of this 
section is to consider these singularities. At a singular point a dyon%
\footnote{By ``dyon'' we mean a state charged electrically or magnetically
or both.}
(or dyons) becomes massless. Therefore, such a point is a transition point 
between the Coulomb branch and the Higgs-confinement branch
\draftnote{ALWAYS? WHAT ABOUT MODELS WITHOUT A HIGGS PHASE?}
and, as such, can be approached from the later \cite{IS9408,EFGR}.

To isolate these points, we perturb the theory by adding to
$\Wtree$ a function of the Coulomb moduli:
\beq\label{Wpert} \Wtree \goto \sum_l \tr(h_l Z_l)+gF(\Phi;\xi). \eeq
(where $g,\xi$ are parameters and for each $\xi$, $F$ is a gauge invariant
function of $\Phi$; we allow a constant term in $F$). As a result, most of 
the Coulomb vacua will be
lifted. In fact, only vacua with massless dyons (that undergo condensation)
can survive such a perturbation \cite{EFGIR}. We will look for such vacua
and verify that at these points (in the limit $g\goto0$) the curve indeed
becomes singular. Classically, these vacua are characterized, among the
generic Coulomb vacua, by an enhanced residual gauge symmetry.
\draftnote{ALSO IN STRONG COUPLING? PROBABLY NOT!}
The perturbation $F$ will be chosen to admit each such vacuum as a solution
(of $dF=0$ and, therefore, also of $d\Wtree=0$, $Q=\tilde{Q}=0$) for some 
value of $\xi$. For all the models considered, a convenient 
perturbation is $F=P(x)|_{x=\xi}$, where $P(x)$ is the generating function for
the gauge invariant moduli (a variant of det$(t-\vev{\Phi})$) -- the polynomial
appearing in the leading term of the curve \cite{EFGIR-proc,EFGIR2}.

The quantum vacua of the
Higgs-confinement branch are the solutions of the equations of motion
$dW_\eff=0$, obtained from the effective superpotential $W_\eff$, so one
way of obtaining them is to find $W_\eff$ and solve the corresponding
equations. However, as considered in \cite{EFGIR}, if one recognizes the low
energy effective model in the vacuum of interest and knows $W_L$ (obtained
from $W_\eff$ by integrating out of massive fields by their equations of
motion but {\em not} taking the limit of infinite masses), it is sometimes
simpler to obtain the vevs of the moduli directly from $W_L$.

We will concentrate on the vacua with $SU(2)\times U(1)^{r-1}$ gauge freedom
(vacua with a larger gauge freedom are always on the boundary of the set of
the $SU(2)$ vacua). Generically, all the $SU(2)$-charged matter fields in these
vacua are massive and can be integrated out, leading to the $SU(2)$ $N=1$
SYM model with two vacua and an effective superpotential
\beq\label{W-SU2-YM} W_d=\pm2\Lm_d^3 \hs. \eeq
Going back up (``integrating in'' \cite{Intriligator}), $W_L$ is related to 
$W_d$ by
\beq\label{W-L}
  W_L=W_d+W_\cl+W_\Dl \hsc
  W_\cl=\Wtree|_{\vph_\cl} \hsc
\eeq
where $\vph$ represents, symbolically, the fields integrated in (in our case
these are all the matter fields -- $Q$ and $\tilde{Q}$ are integrated in 
first and then $\Phi$), $\vph_\cl$ is the classical solution 
($d\Wtree/d\vph|_{\vph_\cl}=0$) and $W_\Dl$ is a
possible additional quantum correction. We will assume $W_\Dl=0$. In all the 
cases considered $W_\cl$ vanishes (because of the constant term in $F$), 
therefore, eq. (\ref{W-L}) reduces to $W_L=W_d$. The inverse Legendre
transform equations of the integrating-in procedure%
\footnote{For more details, see \cite{EFGIR,EFGIR-proc}.}
lead to the following equations
\beq\label{Sing} F|_{\Phi=\vev{\Phi}}=\frac{\pt W_d}{\pt g} \hsc
     g\left.\frac{\pt F}{\pt\xi}\right|_{\Phi=\vev{\Phi}}=
     \frac{\pt W_d}{\pt \xi}.
\eeq
$W_L$ is proportional to $g$ (this follows from the conservation of the $U(1)$
charge $R'$, defined in (\ref{U1})), therefore, eqs. (\ref{Sing}) have the 
form of ``singularity equations'' for the function (of $\xi$)
$F|_{\Phi=\vev{\Phi}}-W_L/g$ (the function and its derivative vanishes).
\draftnote{RANGE OF VALIDITY?}

In the following subsections, we provide the details of the above procedure
for the groups $SU(N_c)$ and $SO(N_c)$.

\subsection{$SU(N_c)$, $N_c>2$}

In this model, symmetry enhancement occurs when some of the $\Phi_a$'s 
coincide. Each such vacuum is a solution of (\ref{Wpert}) with
\beq F(\Phi;\xi)=P|_{x=\xi} \eeq
for some $\xi$, where $P(x)$ is defined in (\ref{SU-P}). When $n$ $\phi_a$'s
coincide, the gauge symmetry is enhanced to
$SU(n)$. In particular,  the $SU(2)$ vacua are
\beq\label{SU-vac} \phi_1=\phi_2=\xi\neq\phi_a \hs \forall a>2. \eeq
The matching of scales is \cite{KSS}%
\footnote{In the determination of the scale matching, we ignore numerical
  factors. These depend on conventions and can be fixed consistently at the 
  end. In the following $\sim$ will denote equality up to such factors.}
\beq \Lm_d^6\sim\frac{\Pf(M_Q)M_\Phi^2}{\prod_3^{N_c}(\xi-\phi_a)^2}\Lm^b, \eeq
where $M_Q$ and $M_\Phi$ are the effective masses of the matter fields charged
under the residual $SU(2)$ and the denominator represents the masses of the
higgsed gluons. $M_Q$ comes from the original superpotential:
\[ \sum_l\tr(h_lZ_l)\goto\tr[h(\xi)(\tilde{q}q)]=
  \half\tr[M_Q({Q'}^T\ep Q')], \]
where
\[ h(\xi)=\sum_l h_l \xi^l \hsc Q'=(q,\ep\tilde{q}^T) \hsc
   \ep=\left(\begin{array}{cc}0 & 1 \\ -1 & 0 \end{array}\right), \]
and $q,\tilde{q}$ consist of the first (color) components of $Q,\tilde{Q}$,
respectively. This gives
\beq M_Q\sim\left(\begin{array}{cc}
   0 & -h(\xi)^T \\ h(\xi) & 0 \end{array}\right), \eeq
which leads to
\beq \Pf(M_Q)\sim\det(h(x))\equiv H(\xi). \eeq

$M_\Phi$ comes from the perturbation $F$ and it is given by \cite{EFGIR,TY}
$M_\Phi=g\prod_3^{N_c}(\xi-\phi_a)$.
Combining all this, we obtain
\[ \Lm_d^6\sim g^2H(\xi)\Lm^b , \]
and eqs. (\ref{Sing}) take the form of singularity equations for the
functions 
\beq P_\pm(x)=P(x)\pm\sqrt{H(x)\Lm^b} \eeq
(where we have chosen a normalization for $\Lm$).
The solutions are naturally also
solutions for the singularity equations for the polynomial $K=P_+P_-$ which
is indeed the polynomial of the curve (\ref{SU-curve}).
\draftnote{WHAT ABOUT THE ADDITIONAL SINGULARITIES $P_+=P_-=0$? (these are
quark singularities)}

\subsection{$SO(N_c)$}

There are two types of symmetry enhancement in this model:
\begin{itemize}
  \item $n$ $\phi_a^2$'s coincide but do not vanish $\goto SU(n)$;
  \item $n$ $\phi_a$'s vanish $\goto SO(2n+\ep)$ (where $N_c=2r+\ep$)
\end{itemize}
and the superpotential (\ref{Wpert}) with
\beq F(\Phi;\xi)=P|_{x=\xi} \hsc P(x)=\prod_a(x-\phi_a^2) \eeq
admits each of these vacua as solutions, for an appropriate $\xi$.
Each of the above types contains $SU(2)$ vacua
(the second one -- $SO(3)$ -- appears only for $N_c$ odd).

The $SU(2)$ vacua of the first type are of the form
\beq\label{SO-vac1}
  \phi_1^2=\phi_2^2=t^2\equiv\xi\neq\phi_a^2 \hs \forall a>2 \hsc
  \phi_a\neq0 \hs \forall a.
\eeq
The matching of scales is \cite{TY}
\beq
  \Lm_d^6\sim\Pf(M_Q)g^2 t^{2(2-\ep)}\Lm^b.
\eeq
$M_Q$ comes from the original superpotential:
\[ \sum_l\tr(h_lZ_l)\goto{Q'}^T U^T U
   \left(\begin{array}{cc}h(-t)^T & 0 \\ 0 & h(t)^T \end{array}\right) Q'
   \hsc U^T U=\left(\begin{array}{cc}0 & 1 \\ 1 & 0 \end{array}\right)
\]
where $U$ is the rotation (\ref{rot}) that diagonalizes $\vev{\phi}$
and  $Q'=(q_1,\ep q_2)$ consists of the first (color) components of
$U^\dagger Q$,
expressed in terms of two $SU(2)$-fundamentals $q_1$,$q_2$. This gives
\[ \sum_l\tr(h_lZ_l)\goto\half\tr[M_Q(q^T\ep q)] \hsc
   M_Q\sim\left(\begin{array}{cc}
   0 & -h(t)^T \\ h(t) & 0 \end{array}\right), \]
(a $4N_f\times4N_f$ matrix) which leads to
\beq \Pf(M_Q)=\det(h(t))\equiv H(\xi). \eeq
The scale matching is, therefore,
\[ \Lm_d^6\sim g^2\xi^{2-\ep}H(\xi)\Lm^b \]
and eqs. (\ref{Sing}) take the form of singularity equations for the
functions
\beq P_\pm(x)\sim P(x)\pm\sqrt{x^{2-\ep}H(x)\Lm^b} \eeq
(for an appropriate normalization choice).
The solutions are naturally also
solutions for the singularity equations for the polynomial $K=xP_+P_-$ which
is indeed the polynomial of the curve (\ref{SO-curve}).
\draftnote{WHAT ABOUT THE ADDITIONAL SINGULARITIES $P_+=P_-=0$ ?}

\subsection{Vacua with Embedding Index = 2}

The $SO(2r+1)\goto SO(3)$ vacua are of the form
\beq\label{SO-vac2}  \phi_1=0\neq\phi_a \hs \forall a>1. \eeq
The embedding index of this subgroup in $SO(2r+1)$ is 2 and the  
matching of scales is
\beq \Lm_d^3\sim\frac{\Pf(M_Q)M_\Phi}{\prod_2^{r}\phi_a^4}\Lm^b, \eeq
where $M_Q$ and $M_\Phi$ are the effective masses of the matter fields charged
under the residual $SO(3)$ ($2N_f+1$ vectors) and the denominator represents
the masses of the higgsed gluons.
In this case, the one instanton contribution is of the same order as the
gaugino condensate and, therefore, must be taken into account; it should
lead to the singularities at $x=s_{2r}=0$.

Similarly, for $USp(2r)$, the $SU(2)$ vacua of the form
\beq\label{Sp-vac1}
  \phi_1^2=\phi_2^2=t^2\equiv\xi\neq\phi_a^2 \hs \forall a>2 \hsc
  \phi_a\neq0 \hs \forall a
\eeq
also have embedding index = 2, and the matching of scales is 
\beq
  \Lm_d^6\sim\Pf(M_Q)\frac{g^2}{t^4}\Lm^{2b} .
\eeq
Again, the one instanton contribution is important in this branch.
In addition, one should consider the $USp(2)$ vacua at $\phi_1=0$;
this will not be done here.

\newsection{Discussion: Non-Trivial Super-Conformal Fixed Points}

For every model discussed in this article, its curve is identical in shape
($x$ and $\phi_a$ dependence) to an $N=2$ model of the same type (gauge group
and matter representations), with a different number of flavors
$N'_f=\deg(H(x))$, where the quark masses (or the squares of the
masses, for $SO(N_c)$ and $USp(N_c)$) are the roots of $H(x)$ and the scale
$\Lm$ is rescaled. For example, for $\det h_{l_\max}\neq0$, one obtains
$N'_f=l_\max N_f$ and ${\Lm'}^{b'}=\Lm^b\det(h_{l_\max})$ (for $USp(N_c)$,
${\Lm'}^{b'}=\Lm^b\sqrt{\det(h_{l_\max})}$). This means that calculations
performed for the $N=2$ models, using only the curve, apply directly also
to the more general models. One implication of this is that
the curve passes automatically some other consistency conditions which were
not discussed here but were checked for the $N=2$ cases in previous works
(\eg the monodromy structure).

Another important implication is that the Coulomb branch of these models has
many points where mutually non-local dyons are massless \cite{AD,APSW,EHIY}
and presumably correspond to non-trivial (interacting) super-conformal field
theories (SCFTs).
Moreover, the authors of \cite{APSW,EHIY} also calculated the critical
exponents of perturbations around these points. The ratios of these
exponents are determined entirely by the curve%
\footnote{To determine the normalization, they used additional $N=2$
information that is not currently available in $N=1$ models.},
therefore, the same ratios appear in the $N=1$ models discussed here,
and these ratios imply that the above distinguished points in the $N=1$
Coulomb phase correspond to many {\em different} SCFTs, corresponding to
the universality classes listed in \cite{EHIY}.

For $l_\max>1$, these SCFTs seem to be different from the corresponding
$N=2$ SCFTs, the difference being in the number of flavors and,
therefore, in the global symmetry (this was observed, for the $SU(N_c)$
case in \cite{Kapustin}). If this is true, these are new examples of
$N=1$ SCFTs. However, this is not the only possibility%
\footnote{We are grateful to M.R. Plesser for a discussion on this
point.}.
Indeed, it may happen that at a conformal point, the global
symmetry and supersymmetry is enhanced. Evidence
for supersymmetry enhancement in some cases appear, for example, in
\cite{LS,CL}.
\draftnote{Examples of global (bosonic) symmetry enhancement are
hep-th/9611020 (SU(N)) and hep-th/9701191 (SO(N)).}
Therefore, the SCFTs described
here may have $N=2$ supersymmetry. In this case, the (ratios of) scaling
dimensions suggest that these are actually the SCFTs of the $N=2$ models
corresponding to the same curve. Yet, another possibility is that these
are {\em new} $N=2$ SCFTs. For example, these may be some variants of the 
``type E'' SCFTs -- new fixed points with exceptional symmetries -- 
considered in \cite{E-SCFT}. In this case, the present model
would provide a field-theoretical description of these SCFTs, which is
currently unknown. Any one of these alternatives is interesting and
invites further study.

\vspace{1cm}
\noindent{\bf Acknowledgment}

\noindent We thank S. Elitzur, A. Forge and R. Plesser for discussions. 
This work is supported in part by BSF -- American-Israel Bi-National
Science Foundation, and by the Israel Science Foundation founded by the 
Israel Academy of Sciences and Humanities -- Centers of Excellence Program.

\vspace{1cm}
\noindent{\bf Note Added}

\noindent As this article was being completed, we received the preprint 
\cite{KTY} which overlaps parts of section 3.


\appendix
\renewcommand{\newsection}[1]{
 \vspace{10mm} \pagebreak[3]
 \addtocounter{section}{1}
 \setcounter{equation}{0}
 \setcounter{subsection}{0}
 \setcounter{paragraph}{0}
 \setcounter{equation}{0}
 \setcounter{figure}{0}
 \setcounter{table}{0}
 \addcontentsline{toc}{section}{
  Appendix \protect\numberline{\Alph{section}}{#1}}
 \begin{flushleft}
  {\large\bf Appendix \thesection. \hspace{5mm} #1}
 \end{flushleft}
 \nopagebreak}


\newsection{The One Instanton Factor}
The one-instanton factor in an asymptotically free model is $\Lm^b$, 
where $\Lm$ is the dynamically generated scale and $b$ is proportional to
the one-loop
coefficient of the gauge coupling beta function. For an $N=1$ supersymmetric
model with matter (hypermultiplets) in (irreducible) representations $R_i$ of 
the gauge group, $b$ is given by $b=3k_A-\sum k_i$, where $k_i$ is the Dynkin 
index of the representation $R_i$ and, in particular, $k_A$ is the index of 
the adjoint representation.
We summarize in this Appendix the Dynkin indices needed for the calculation of
the one-instanton factors for the models considered in the text:
\beq
 \begin{array}{|c||c|c|c|}\hline
          & k_f   & k_A & b=2k_A-2N_fk_f \\ \hline
  SU(N_c) & \half & N_c & 2N_c-N_f    \\
  SO(N_c) &   1   & N_c-2 & 2(N_c-2-N_f)   \\
  USp(2r) & \half & r+1   & 2r+2-N_f   \\ \hline
 \end{array}
\eeq

\newsection{Calculation of Discriminants}
In this Appendix we present the calculation of discriminants used in the text
(extending an approach used in \cite{HO}).
We begin with some generalities (more information can be found in\cite{Lang}).
The {\em discriminant} of a polynomial
\[ P(x)=\sum_{i=0}^r P_i x^{r-i}=P_0\prod_{a=1}^r(x-p_a) \]
is defined to be
\beq \Dl_P=P_0^{2r-1}\prod_{a<b}(p_a-p_b)^2. \eeq
It is useful to define also the {\em resultant} of two polynomials
\beq \Dl(P,Q)=P_0^s \prod_{a=1}^rQ(p_a)=P_0^s Q_0^r \prod_{a,b}(p_a-q_b) \eeq
(where $s$ is the degree of $Q$). Then the discriminant of the product $PQ$
is%
\footnote{In this Appendix ``$\sim$'' will mean ``equality up to a
  non-zero numerical constant'', \eg a sign or a power of $P_0$.}
\beq\label{Dl-Prd} \Dl_{PQ}\sim\Dl_P\Dl_Q\Dl(P,Q)^2. \eeq
Also, since $P'(p_a)=P_0\prod_{b\neq a}(p_a-p_b)$,
\beq\label{Dl-Der} \Dl_P\sim\Dl(P,P')\sim\prod_{b=1}^{r-1}P(q_b), \eeq
where $\{q_b\}$ are the roots of $Q=P'$
(note that the resultant is symmetric in its arguments, up to a sign).

In the following we consider a polynomial
\beq K(x;\ep)=\sum_\al \ep^\al K_\al(x) \hs, \eeq
where $\ep$ is a small (non-zero) parameter and $\deg(K)=\deg(K_0)$.
We evaluate the discriminant of $K$ in the leading order in $\ep$. First
we summarize the results:

\begin{itemize}
  \item{$K_0=P^2$}: ($P(x)=\prod_{a=1}^r(x-p_a)$, $\Dl_P\neq0$)
    \beq\label{Dl-PP}
      \Dl_K\sim(\Dl_P)^2\prod_{a=1}^r K(p'_a)(1+o(\ep^0))\hs,
    \eeq
    where
    \beq\label{Kp1-PP}
      K(p'_a)=\ep K_1(p_a)+o(\ep)
    \eeq
    and if $K_1(p_a)=0$ then
    \beq\label{Kp0-PP}
      K(p'_a)=\ep^2 \left[K_2-\left(\frac{K_1}{2P}\right)^2\right]_{x=p_a}
             +o(\ep^2)\hs.
    \eeq

  \item{$K_0=xP^2$}:
    ($P$ as above and also $P(0)\neq0$)
    \beq\label{Dl-xPP}
      \Dl_K\sim(\Dl_P)^2 P(0)^3\prod_{a=1}^r K(p'_a)(1+o(\ep^0))\hs,
    \eeq
    where
    \beq\label{Kp1-xPP}
      K(p'_a)=\ep K_1(p_a)+o(\ep)
    \eeq
    and if $K_1(p_a)=0$ then
    \beq\label{Kp0-xPP}
      K(p'_a)=\ep^2 \left[K_2-\frac{1}{x}
              \left(\frac{K_1}{2P}\right)^2\right]_{x=p_a}+o(\ep^2)\hs.
    \eeq

\end{itemize}

\subsection{$K_0=P^2$}
Using (\ref{Dl-Der}) we obtain
\beq \Dl_K\sim\prod_c K(x_c)\hs, \eeq
where $\{x_c\}$ are the roots of $K'$. Since $K'=2PP'+O(\ep)$,
\[
  \{x_c\}=\{p'_a\}\cup\{q'_b\}
  \hsc p'_a=p_a+o(\ep^0)  \hsc q'_b=q_b+o(\ep^0)\hs,
\]
where $\{p'_a\}$ are the roots of $P$ and $\{q'_b\}$ are the roots of $P'$.
Using (\ref{Dl-Der}) again, we obtain
\beq
  \prod_b K(q'_b)\sim[\prod_b P(q_b)]^2+o(\ep^0)
                 \sim(\Dl_P)^2+o(\ep^0)\hs,
\eeq
which leads to (\ref{Dl-PP}).
To estimate $K(p'_a)$ we observe that $x=p'_a$ is a solution of
\[ 0=K'(x)=(2R(p_a)^2+o(\ep^0))(x-p_a)+\ep(K'_1(p_a)+o(\ep^0)) \]
(where $P(x)=R(x)(x-p_a)$ and $R(p_a)\neq0$ since $\Dl_P\neq0$), therefore,
\[ (x-p_a)=\Oc(\ep) \]
and this implies (\ref{Kp1-PP}). If $K_1(p_a)=0$, eq. (\ref{Kp1-PP}) is not
useful and we must refine the estimation. Denoting $K_1(x)=S(x)(x-p_a)^m$,
$x=p'_a$ is a solution of
\[ 0=K'(x)=(2R(p_a)^2+o(\ep^0))(x-p_a)
          +\ep(x-p_a)^{m-1}(mS(p_a)+o(\ep^0))
          +\ep^2(K'_2(p_a)+o(\ep^0)),                \]
therefore,
\beq (x-p_a)=\left\{\begin{array}{ll}
   -\ep\left.\frac{S}{2R^2}\right|_{x=p_a}+o(\ep) & m=1 \\
   -\ep^2\left.\frac{K'_2}{2R^2}\right|_{x=p_a}+o(\ep^2) & m>1
   \end{array}\right.                                          \eeq
and this leads to (\ref{Kp0-PP}).

\subsection{$K_0=xP^2$}

\subsubsection{$K(0)=0$}

In this case, $K=x\tilde{K}$ and $\tilde{K}_0=P^2$, therefore we can use
the results of the previous subsection:
\beq\label{Dl-Kt}
  \Dl_{\tilde{K}}\sim(\Dl_P)^2\prod_a\tilde{K}(p'_a)(1+o(\ep^0))\hs,
\eeq
where
\beq\label{Kp1-Kt}
  \tilde{K}(p'_a)=\ep\tilde{K}_1(p_a)+o(\ep)
\eeq
and for $\tilde{K}_1(p_a)=0$
\beq\label{Kp0-Kt}
  \tilde{K}(p'_a)=\ep^2 \left(\tilde{K}_2
  -\left(\frac{\tilde{K}_1}{2P}\right)^2\right)_{x=p_a}+o(\ep^2)\hs.
\eeq
Multiplying (\ref{Kp1-Kt}) and (\ref{Kp0-Kt}) by $p'_a=p_a+o(\ep^0)$ gives
(\ref{Kp1-xPP}) and (\ref{Kp0-xPP}) respectively. Substituting (\ref{Dl-Kt})
in
\[ \Dl_K=\Dl_{\tilde{K}}\tilde{K}(0)^2
        =\Dl_{\tilde{K}}(P(0)^4+o(\ep^0)) \]
and observing that
\[ \prod_a K(p'_a)=(P(0)+o(\ep^0))\prod_a\tilde{K}(p'_a)\hs, \]
one obtains (\ref{Dl-xPP}).

\subsubsection{$K(0)\neq0$}

In this case $\tilde{K}=xK$ and $\tilde{P}=xP$ obey
$\tilde{K}_0=\tilde{P}^2$ and we can again use the results of the previous
subsection:
\beq\label{Dl-Ktt}
  \Dl_{\tilde{K}}\sim(\Dl_{\tilde{P}})^2\tilde{K}(p'_0)
  \prod_a\tilde{K}(p'_a)(1+o(\ep^0)\hs,
\eeq
where
\beq\label{Kp1-Ktt}
  \tilde{K}(p'_a)=\ep\tilde{K}_1(p_a)+o(\ep)
\eeq
and for $\tilde{K}_1(p_a)=0$,
\beq\label{Kp0-Ktt}
  \tilde{K}(p'_a)=\ep^2 \left(\tilde{K}_2
  -\left(\frac{\tilde{K}_1}{2\tilde{P}}\right)^2\right)_{x=p_a}+o(\ep^2)\hs.
\eeq
$x=p'_0=o(\ep^0)$ satisfies
\beq\label{xKp} 0=(xK)'=(2P(0)^2+o(\ep^0))x+\ep(xQ)'\hs, \eeq
where we use the notation $K=P^2+\ep Q$. As $\ep\goto0$,
$(xQ)'_{x=p'_0}\sim\ep^\bt$ for some non-negative $\bt$, so (\ref{xKp})
implies $p'_0\sim\ep^{\bt+1}$. The terms in $(xQ)'_{x=p'_0}$ are of the form
\[ (xQ)'_{x=p'_0}\sim\ep^\al(p'_0)^k\sim\ep^{\al+(\bt+1)k} \hsc
   k,\al,\bt\geq0  \]
so for the leading term we must have $\al+(\bt+1)k=\bt$. But this is possible
only for $k=0$ (which also implies $\al=\bt$). Therefore,
\[ (xQ)'_{x=p'_0}=Q(0)(1+o(\ep^0)) \]
and (\ref{xKp}) leads to
\beq\label{p0} p'_0=-\frac{K(0)}{2P(0)^2}(1+o(\ep^0))\hs. \eeq
{}From (\ref{Dl-Ktt}) we obtain
\[ K(0)^2\Dl_K=\Dl_{\tilde{K}}
              \sim(\Dl_P)^2 P(0)^5 p'_0 K(0)\prod_a K(p'_a)(1+o(\ep^0)\hs,  \]
which leads, after substituting (\ref{p0}), to (\ref{Dl-xPP}).
Dividing (\ref{Kp1-Ktt}) and (\ref{Kp0-Ktt}) by $p'_a=p_a+o(\ep^0)$ gives
(\ref{Kp1-xPP}) and (\ref{Kp0-xPP}) respectively.
\ifdraft\begindraft
 \input{coulsup}
\enddraft\fi


\end{document}